# Reply to the Comment by A. N. Vaidya and R. de L. Rodrigues on "Solution of the Relativistic Dirac-Morse Problem"

A. D. Alhaidari

Physics Department, King Fahd University of Petroleum & Minerals, Box 5047,
Dhahran 31261, Saudi Arabia
E-mail: **haidari@mailaps.org**

Gauge invariance <u>does not</u> unify the Dirac-Coulomb, Dirac-Oscillator, and Dirac-Morse problems nor does gauge transformation map one into the other. <u>NO</u> claim to the contrary has been made anywhere in our Letter [1]. What we do is that for a <u>given</u> physical problem (e.g. the Dirac-Morse potential) we impose a constraint on the potential components in such away that makes the solution of the wave equation easily attainable. It is specifically stated that our use of gauge invariance was merely to establish that in the case of static and spherically symmetric charge distribution the 4-component electromagnetic potential $[A_0, \vec{A}]$ takes the form $[V(r), \hat{r}W(r)]$. One of the original contributions in our Letter is the choice of constraint which results in Schrödinger-like differential equation for the upper spinor component. This makes the solution of the relativistic problem easily obtainable by correspondence with well-known exactly solvable nonrelativistic problems. Table I in the Letter lists the potential function, physical parameters, and energy spectrum for each of the three problems. Nowhere did we give any transformation (let alone gauge transformation) that maps one set of these quantities into the other. In fact, in the Letter we state that we "choose" or "consider" an expression for the potential function. Of course, we were guided in the case of the Dirac-Coulomb and Dirac-Oscillator problems by well-established results.

On the other hand, it is true that the Hamiltonian which results in the radial Eq. (1) is not the minimum coupling Hamiltonian *H* shown on the first page of the Letter but the one obtained from it by replacing the two off-diagonal terms $\alpha \vec{\sigma} \cdot \vec{A}$ with $\pm i\alpha \vec{\sigma} \cdot \vec{A}$, respectively. This makes our Hamiltonian equivalent to that given by equation (3) in the Comment [2] for $(A_0, \vec{A}) = (V, \hat{r}W)$. Consequently, our interpretation of $(V, \hat{r}W)$ as the electromagnetic potential and our statement that "$W(r)$ is a gauge potential" are not correct. Likewise, calling Eq. (2) in the Letter, or any other derived from it, as the "gauge fixing condition" is not accurate. This has to be replaced everywhere by the term "constraint". Nonetheless, all developments based on and subsequent to Eq. (1) still stand independent of that interpretation.

Dismissing this wrong interpretation we find that the same formalism, which was developed in the Letter starting with Eq. (1), has been applied successfully to another class of relativistic shape invariant potentials [3]. This class includes "Dirac-Rosen-Morse", "Dirac-Eckart", "Dirac-Pöschl-Teller", and "Dirac-Scarf" potentials. These considerations give further support to the method used and to the results thus obtained in the Letter.

In nonrelativistic quantum mechanics the relationship among the Oscillator, Coulomb, and Morse problems is well established by many authors over the years [4].

Each one of these problems, including its energy spectrum and wave functions, could be mapped into the other by point canonical transformations [5]. The relativistic extension of this relationship has been established only recently [6] where the relativistic version of these three problems are also interrelated and belong to the same class which carries a representation of a superalgebra which is a graded extension of so(2,1) Lie algebra. Each of these relativistic problems can be mapped into one another by an "extended point canonical transformation" – not a gauge transformation. The results of that development fully agree with those in our Letter giving another independent method for proving the validity of our established findings. Therefore, the conclusion reached by A. N. Vaidya and R. de L. Rodrigues in their Comment is not supported.